\documentclass[aps, prd, a4paper, 10pt, showpacs, superscriptaddress, floatfix, nofootinbib, reprint]{revtex4-1}
\usepackage{color}
\usepackage{xcolor}
\usepackage{amsmath,amssymb,graphicx}
\usepackage{gensymb}
\usepackage{wasysym}
\usepackage{bm,lipsum}
\usepackage{times}
\usepackage{multirow}
\usepackage{microtype}
\usepackage{comment}
\usepackage[utf8]{inputenc}
\usepackage{booktabs}
\usepackage{subfigure}
\usepackage[normalem]{ulem}
\usepackage[varg]{txfonts}
\usepackage{bm,graphics,graphicx,epsfig,color,ulem}
\usepackage{array}
\usepackage[colorlinks, pdfborder={0 0 0}]{hyperref}
\definecolor{LinkColor}{rgb}{0.75, 0, 0}
\definecolor{CiteColor}{rgb}{0, 0.5, 0.5}
\definecolor{UrlColor}{rgb}{0, 0, 0.75}
\hypersetup{linkcolor=LinkColor}
\hypersetup{citecolor=CiteColor}
\hypersetup{urlcolor=UrlColor}
\allowdisplaybreaks
\newcolumntype{C}{>{\arraybackslash}m{6cm}}

\def\hbbh{{\em h}BBH}
\usepackage{tabularx}

\begin{document}

\title{Archival searches for stellar-mass binary black holes in {\em LISA}}
\author{Becca Ewing}
\affiliation{Institute for Gravitation and the Cosmos, Department of Physics, Pennsylvania State University, University Park, PA, 16802, USA}

\author{Surabhi Sachdev}
\affiliation{Institute for Gravitation and the Cosmos, Department of Physics, Pennsylvania State University, University Park, PA, 16802, USA}

\author{Ssohrab Borhanian}
\affiliation{Institute for Gravitation and the Cosmos, Department of Physics, Pennsylvania State University, University Park, PA, 16802, USA}

\author{B. S. Sathyaprakash}
\affiliation{Institute for Gravitation and the Cosmos, Department of Physics, Pennsylvania State University, University Park, PA, 16802, USA}
\affiliation{Department of Astronomy \& Astrophysics, Pennsylvania State University, University Park, PA, 16802, USA}
\affiliation{School of Physics and Astronomy, Cardiff University, Cardiff, UK, CF24 3AA}

\begin{abstract}
Stellar-mass binary black holes will sweep through the frequency band of the Laser Interferometer Space Antenna ({\em LISA}) for months to years before appearing in the audio-band of ground-based gravitational-wave detectors. One can expect several tens of these events up to a distance of $500 \,\mathrm{Mpc}$ each year. The {\em LISA} signal-to-noise ratio for such sources even at these close distances will be too small for a blind search to confidently detect them. However, next generation ground-based gravitational-wave detectors, expected to be operational at the time of {\em LISA}, will observe them with signal-to-noise ratios of several thousands and measure their parameters very accurately. We show that such high fidelity observations of these sources by ground-based detectors help in archival searches to dig tens of signals out of {\em LISA} data each year. 

\end{abstract}

\date{\today}
\maketitle

\section{Introduction and background}
The discovery of GW150914~\cite{Abbott:2016blz} by the Advanced Laser Interferometer Gravitational-wave Observatory (LIGO)~\cite{TheLIGOScientific:2014jea} and the continued detections of stellar mass binary black holes~\cite{TheLIGOScientific:2016pea, LIGOScientific:2018mvr, Abbott:2020niy} by Advanced LIGO and Virgo~\cite{TheVirgo:2014hva} set the stage for observing such systems with the Laser Interferometer Space Antenna ({\em LISA}) \cite{Sesana:2016ljz} when it comes online. GW150914 is the result of the merger of a pair of $36^{+5}_{-3}\,M_\odot$  and $31^{+3}_{-4}\,M_\odot$ black holes at a distance of $440^{+150}_{-170}\,\rm  Mpc$ \cite{LIGOScientific:2018mvr}. The companion masses are larger than what was initially thought possible from stellar evolution \cite{Fryer:2003} (see, however, \cite{Belczynski:2010tb}). It therefore earned the adjective {\em heavy} for black holes in the mass range $\sim [20\,M_\odot, 100\,M_\odot]$ now routinely observed by LIGO and Virgo \cite{Abbott:2017vtc, Chatziioannou:2019dsz, LIGOScientific:2018mvr, Nitz:2019hdf, Venumadhav:2019tad, Abbott:2020niy}. Such heavy binary black holes (\hbbh s) within $\sim1$ Gpc could also be visible in the {\em LISA} band \cite{Sesana:2016ljz} at an earlier stage in their evolution, albeit with a signal-to-noise ratio (SNR) of a few. 

The search for stellar-mass binary black holes in {\em LISA} data could take formidable computational resources \cite{Moore:2019pke}. The resulting false alarm rate due to the large number of templates~\cite{Dhurandhar:1992mw} required would mean that only a handful of nearby sources with SNRs greater than 14 might be detected for a $p$-value of $10^{-3}$. Note, however, that third generation (3G) ground-based observatories, such as the Einstein Telescope \cite{Punturo:2010zz,Hild:2010id} and the Cosmic Explorer \cite{Evans:2016mbw}, operating at the same time as {\em LISA} would observe these sources some months to years after the signal passes the {\em LISA} band, with far greater SNRs compared to those in {\em LISA} and determine the source parameters to a good accuracy. Narrowing down the source parameters by ground-based detectors should then help in archival searches for such systems in {\em LISA} data by reducing the parameter space and hence, false alarm rates and computational costs. Current estimates still require an SNR threshold of $\sim 8$ \cite{Wong:2018uwb} to $\sim 14$ \cite{Moore:2019pke}. 

Multiband observations of \hbbh\ systems in {\em LISA} and ground-based detectors would greatly benefit the science return of these observatories \cite{Vitale:2016rfr,  Barausse:2016eii,  Cutler:2019krq, Carson:2019kkh,  Gupta:2020lxa,  Datta:2020vcj,  Gerosa:2019dbe,  Jani:2019ffg,  Grimm:2020ivq,  Ng:2020jqd,  Liu:2020nwz}. This is because the parameter degeneracies that are present in the later part of the system's evolution in ground-based detectors could be resolved by observing the earlier part of the system's evolution in {\em LISA}. Several authors \cite{Barausse:2016eii,Cutler:2019krq,Carson:2019kkh,Gupta:2020lxa,Datta:2020vcj} have demonstrated that this can principally yield tests of general relativity orders of magnitude better than what would be possible with either detector or detector-network by itself.  What is critical to making that science possible is to unambiguously detect the signals in the {\em LISA} band.

In this paper we will show that 3G observatories will pin down the parameters of \hbbh\ systems well enough to reduce the number of templates required for matched-filter searches to detect such systems in {\em LISA} data to a mere $\mbox{few }\times 10^4$ as opposed to previous estimates of $\sim 10^{12}.$ This means that it will be possible to identify \hbbh\ signals in {\em LISA} data with an SNR of 4 or more with a $p$-value of $10^{-2}$ or better. This will increase the number of sources that will be available for joint observation by both space-borne and ground-based observatories and hence enhance the science return of multiband observations.

The rest of the paper is organized as follows: In Sec.\ \ref{sec:3g}, we will compute for the joint \hbbh\ population expected to be observed the visibility and measurement capabilities of 3G observatories. We will discuss, in particular, the uncertainties in the sky localizations, masses, and spins of the companion black holes---parameters that would need to be searched for in {\em LISA} data. In Sec.\ \ref{sec:mapping}, we show how the problem of assessing {\em LISA}'s performance in observing binary black holes can be mapped to the audio-frequency band. This is possible since there is no mass scale in general relativity: waveforms from binary black holes of different total mass will all look \emph{exactly} the same as long as all other parameters remain the same except for a rescaling of time. This helps in using tools that have been developed for the analysis of ground-based detectors such as the LSC Algorithm Library \cite{lalsuite}. In Sec.\ \ref{sec:archival searches}, we will estimate the number of templates required to search for \hbbh\ systems in {\em LISA} data using the knowledge of parameter accuracies from 3G observatories. We will use two complementary methods to compute the number of templates. The first method works out the invariant volume of the signal manifold over the relevant range of parameters and then divides it by the fraction of volume covered by each template. This gives the minimal number of templates required for archival searches. In a realist data analysis pipeline, however, one needs to make a choice of templates based on a template placement algorithm \cite{Owen:1998dk}. We will use one such algorithm \cite{Babak:2006ty} to get a more realistic estimate of the number of templates. In Sec.\ \ref{sec:sims}, we characterize the efficiency of the template bank by computing the overlap of \hbbh\ waveforms with random parameters maximized over the set of templates in the template bank. We will also discuss in Sec.\ \ref{sec:visibility lisa} the distribution of the SNRs of the sub-population of sources that will be observed by both {\em LISA} and 3G observatories. We conclude in Sec.\,\ref{sec:conclusions} with a summary of the results and future plans.

\section{Stellar Mass Binary Black Holes in 3G Observatories} \label{sec:3g}

3G detectors like Cosmic Explorer (CE) and the Einstein Telescope (ET) will observe stellar-mass binary black holes all the way up to redshifts of $\sim$ 10--50 \cite{Sathyaprakash:2012jk, Hall:2019xmm, Vitale:2016icu}, depending on the intrinsic parameters of the source such as its masses and spins, as well as extrinsic parameters such as the position of the source on the sky and the orientation of the binary's orbit relative to the observer's line of sight. {\em LISA} could observe a small fraction of such systems if they are within $\sim 1\, \rm Gpc$ \cite{Sesana:2016ljz}, but digging them out of the background noise in a blind search will take formidably large computational resources due to the large number of matched filters needed to cover the full parameter space of masses, spins, and sky position \cite{Moore:2019pke}. CE and ET will observe binary black holes that are this close with SNRs of several hundreds to several thousands and determine their parameters with extremely good precision. Such high-fidelity observations will narrow down the search space in the {\em LISA} frequency band, which greatly reduces both the computational resources required but also the background noise from the large number of templates needed in a blind search. 

In this Section we will begin with the visibility of stellar-mass binary black holes in ground-based detectors and then go on to describe the precision with which the parameters can be measured. We shall show that all of the parameters but the chirp mass will be measured by a network of 3G observatories with an accuracy better than {\em LISA} which implies that it will only be necessary to construct templates in chirp mass for LISA.

\subsection{Visibility} \label{sec:3g visibility}

\begin{figure*}
    \centering
    \includegraphics[width=0.49\textwidth]{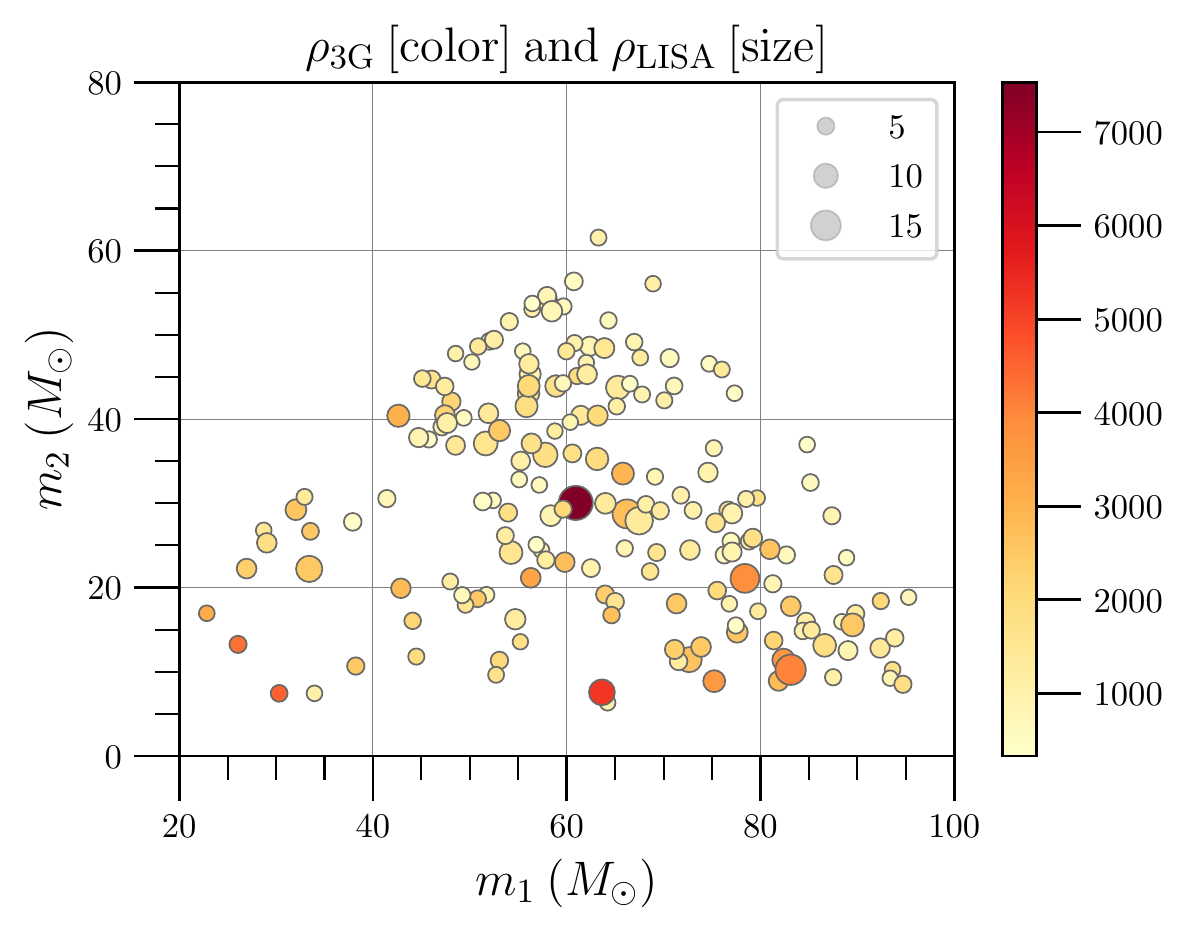}
    \includegraphics[width=0.49\textwidth]{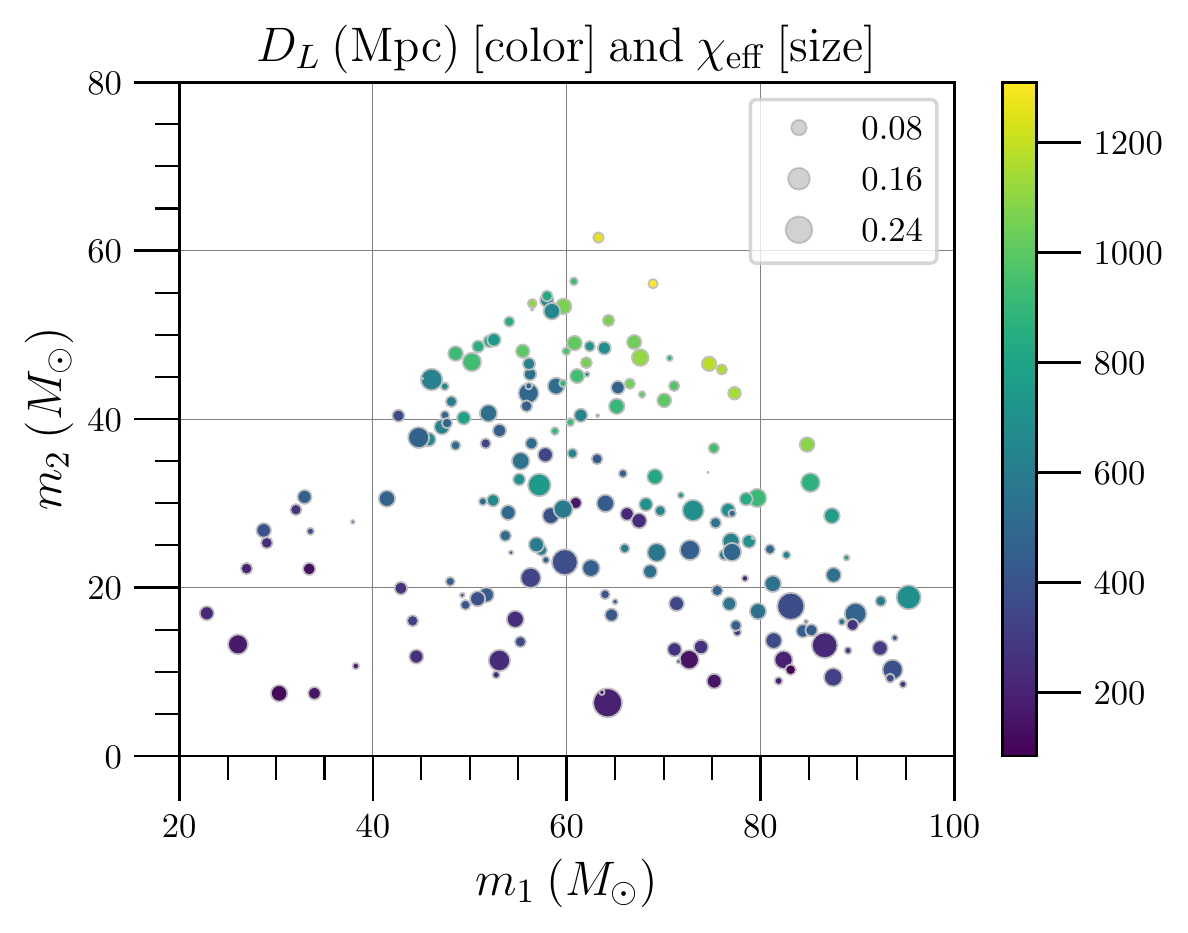}
    \caption{The plot shows the distribution of the 181 \hbbh{s} with a {\em LISA} SNR $\rho_{\it LISA}\ge4$, the two axes show the masses of the companions in both panels. In the left plot, color bar is the 3G SNR $\rho_{\rm 3G}$ and the size of circles depict the {\em LISA} SNR $\rho_{\it LISA}$; in the right plot, color bar is the luminosity distance $D_L$ and the size of the circles represent the effective spin parameter $\chi_{\rm eff}$.}
    \label{fig:visibility}
\end{figure*}

In order to assess what systems will be observable by both {\em LISA} and a network of 3G detectors, we simulated a population of $5\times10^5$ \hbbh{s} which are uniformly distributed in co-moving volume up to a redshift of $z=10$. The companion masses are chosen to follow a power law~\cite{1955ApJ...121..161S} for the larger companion $m_1\in[5 M_{\odot},100M_\odot]$, $p(m_1)\propto m_1^{-\alpha}$ with exponent $\alpha=1.6$, and a uniform distribution in $[5 M_{\odot},m_1]$ for the lighter companion $m_2$ \cite{LIGOScientific:2018jsj}. The companion spins are assumed to be aligned or anti-aligned with the orbital angular momentum, and are drawn from a Gaussian distribution with 0 mean and a standard deviation of 0.1.

We calculated the SNR, $\rho_{\it LISA}$, in {\em LISA} with the estimated power spectral density (PSD) provided in Ref.\,\cite{Cornish:2018dyw} and by marginalizing over the angular dependencies of the signal. We found 181 of the simulated signals to be visible in {\em LISA} with $\rho_{\it LISA}\ge4$ and used these systems as candidates for our 3G-assisted archival search study for {\em LISA} data \cite{Gupta:2020lxa}. We will justify this choice of SNR in Sec.\,\ref{sec:visibility lisa}.  Our choice for a 3G network consists of one ET and two CEs located at fiducial sites in Cascina (Italy), Idaho (USA), and New South Wales (Australia), respectively. The detector sensitivities are set to ET-D for the ET detector and CE1 ($40\,\mathrm{km}$, compact-binary optimized) for the two CEs \cite{Borhanian:2020ypi}.

The left panel of Fig.\,\ref{fig:visibility} shows the distribution of the 3G network SNR $\rho_{\rm 3G}$ and $\rho_{\it LISA}$ against the companion masses of the binaries for the 181 systems. All signals will be detected with SNRs of order $\sim1000$ in 3G, a few reaching values almost 10 times as large. The left panel also confirms the expectation that loud \hbbh{} events in the {\em LISA} band produce loud signals in CE and ET detectors. The right panel presents in a similar fashion the distribution of the luminosity distance, $D_L$, and effective spin parameter, $\chi_{\rm eff}=(m_1\chi_1+m_2\chi_2)/M$~\cite{Racine:2008qv, Ajith:2009bn}, indicating that most of the systems are found at luminosity distances $\lesssim 1\,\mathrm{Gpc}$.

We draw attention, in particular, to the visibility of \hbbh{} in {\em LISA}. The rate for \hbbh{} systems is constrained by the rate of BBH mergers whose up-to-date value is $R=23.8^{+14}_{-8.7}\,{\rm Gpc}^{-3}\, {\rm yr}^{-1}$ \cite{Abbott:2020gyp, Abbott:2020niy}.  Thus, if we take into account that only a fraction $f$ of the binaries contain at least one heavy black hole ($>20 M_\odot$), where
\begin{equation}
f \simeq 1.9 \int_{20\,M_\odot}^{100\,M_\odot}\mathrm{d}m_1\,m_1^{-1.6}\int_{5\,M_\odot}^{m1}\mathrm{d}m_2 \frac{1}{m_1-5\,M_\odot}\simeq 0.32,   
\end{equation}
and 1.9 is the normalization factor, we obtain a median rate $R_{\text{\hbbh}}=f\,R=7.6\,{\rm Gpc}^{-3}\, {\rm yr}^{-1}$ for \hbbh{} merger. Although heavier binaries can be seen to a greater distance their relatively lower prevalence means that it is more likely that we will observe lighter binaries in {\em LISA} more frequently. The detection of lighter binaries in {\em LISA} is also aided by the large SNRs and high fidelity measurements of the 3G network. We also note that the \hbbh{} systems are likely to have larger mass ratios, which is due to the power-law distribution of the primary companion and flat-distribution of the secondary. The large mass ratio is also responsible for low effective spins of \hbbh{} systems as seen in the right panel of Fig.\,\ref{fig:visibility}.

\subsection{Measurability} \label{sec:3g measurability}

\begin{figure*}
    \centering
    \includegraphics[width=1.0\textwidth]{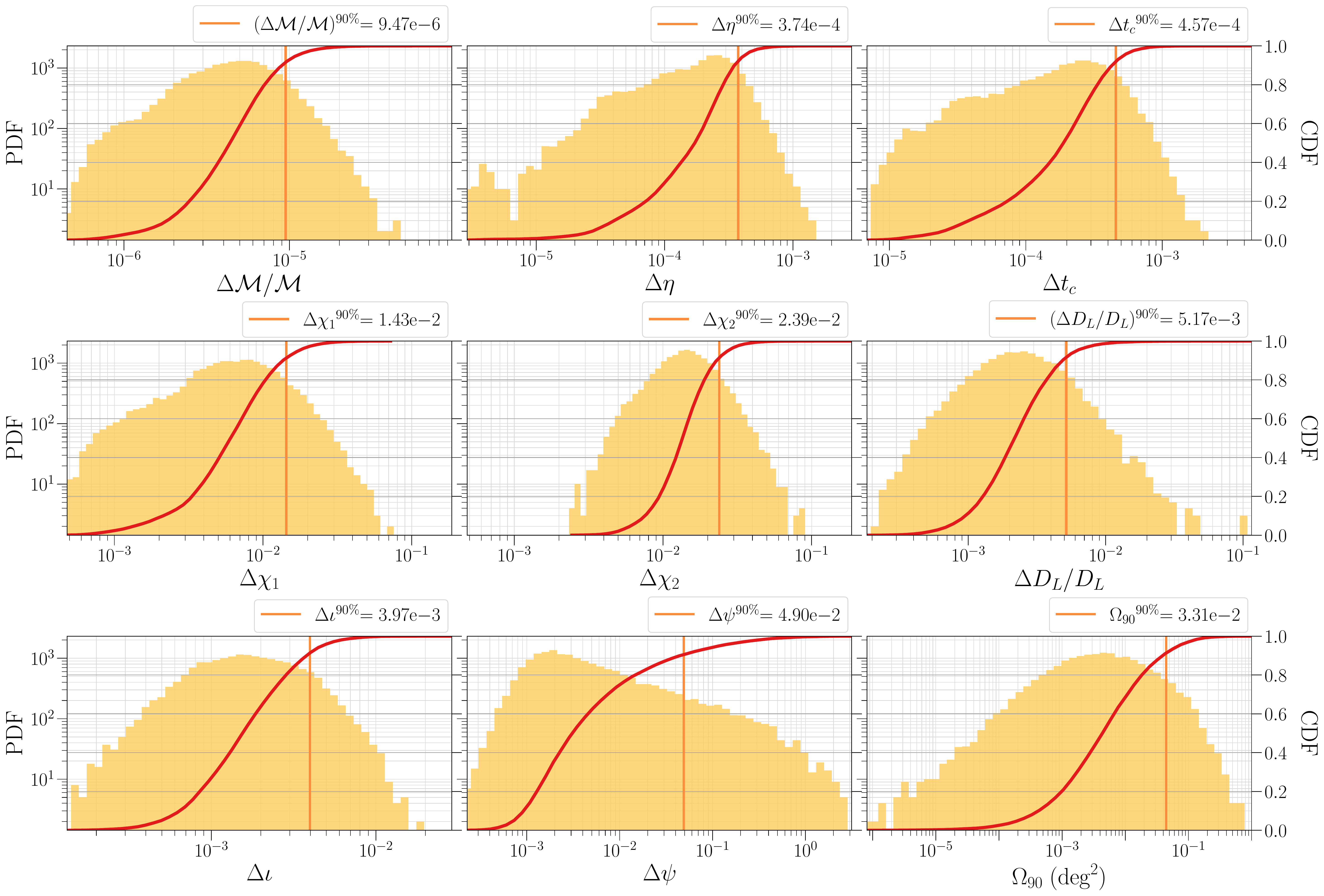}
    \caption{The plot shows the probability and cumulative density functions of the measurement error for the nine parameters of interest for the 181 binary black hole mergers that can be detected in the archival search of {\em LISA} data after they are identified and measured in the 3G data. We simulated 100 realizations for the sky location and binary orientation angles to show the range of parameter errors to be expected from 3G networks.}
    \label{fig:measureability}
\end{figure*}

Next, we want to assess the quality of the parameter estimation that a 3G network can achieve. This crucial information allows us to perform the archival search more efficiently by decreasing the dimension and volume of the parameter space for a template search.

We perform this assessment with \texttt{gwbench} \cite{Borhanian:2020ypi}, a Python package that implements the well-known Fisher information \cite{Cutler:1994ys,Poisson:1995ef,Balasubramanian:1995bm} formalism, and estimate the $1\sigma$-error bounds for each of the 181 systems. The formalism provides an analytical approximation of the Gaussian noise likelihood around its maximum and thus allows us to estimate the measurement errors $\sigma_{\lambda_i} = \sqrt{\Sigma_{ii}}$ on a set of parameters $\bm{\lambda}$ from the covariance matrix $\Sigma$ in the likelihood:

\begin{equation}
    P(\bm{\lambda}) \sim
    \mathrm{e}^{-\frac{1}{2}\,\Sigma^{-1}_{ij}\,\Delta\bm{\lambda}_i\Delta\bm{\lambda}_j}.
\end{equation}
Given a model for the detector response $\tilde{h}(f;\bm{\lambda})$, we can calculate the covariance matrix as the inverse of the Fisher information matrix $\Gamma$

\begin{equation}
    \Sigma^{-1}_{ij} \equiv \Gamma_{ij} = \langle \partial_{\lambda_i} \tilde{h}(f;\boldsymbol{\lambda}), \partial_{\lambda_j} \tilde{h}(f;\boldsymbol{\lambda}) \rangle. \label{eq:fisher}
\end{equation}
The scalar product between two waveforms $\tilde{h}(f;\boldsymbol{\lambda}_1)$ and $\tilde{g}(f;\boldsymbol{\lambda}_2)$ is defined as

\begin{equation}
   \langle \tilde{h}(\boldsymbol{\lambda}_1), \tilde{g}(\boldsymbol{\lambda}_2) \rangle 
   = 2 \int_0^\infty \frac{\tilde{h}(f,\boldsymbol{\lambda}_1)\,\tilde{g}^*(f,\boldsymbol{\lambda}_2) + c.c.}{S_h(f)}\,\mathrm{d}f.
    \label{eq:scalar_product}
\end{equation}
where $\tilde{g}^*$ denotes the complex conjugate of $\tilde{g}$. Note that although the limits in the integral range from 0 to $\infty$, in reality the detector noise power spectral density $S_h(f)$ outweighs the signal power outside a finite frequency range $[f_1, f_2]$ and often the waveform itself will have no support above a frequency $f_{\rm cut} = f(\boldsymbol{\lambda})$ determined by its intrinsic parameters. Thus, the integral gets most of its support over a finite range of frequency $[f_1,f_2]$.

The error bounds that a network of several detectors can achieve are readily computed via the network Fisher matrix 

\begin{equation}
    \Gamma_{\rm net} = \sum_{d} \Gamma_d,
\end{equation}

\noindent which is the sum of the Fisher matrices $\Gamma_d$ for all the detectors in the network. Hence, given a detector response model, we calculate all the detector Fisher matrices and invert their sum to obtain the network covariance matrix from which we extract the desired error bounds.

Lastly, we perform two sanity checks to avoid including faulty numerical data. We first disregard any Fisher matrix $\Gamma$ whose condition number $c_\Gamma = e_M/e_m$ exceeds a threshold value of $10^{15}$ to avoid inverting matrices that are ill-conditioned for this numerical task. $e_M$ and $e_m$ are the maximum and minimum eigenvalues of $\Gamma$. Further, we scrutinize all inversions, if any calculated error bound is smaller than the inversion error $\epsilon=||\Sigma\cdot\Gamma-I||_{\rm max}$. Here, $I$ and $||\cdot||_\text{\rm max}$ represent the identity and maximum matrix norm, respectively.

The loudness of the \hbbh{} signals in CE and ET detectors allows us to make use of waveform models that include higher-order spherical harmonic modes which capture more physical information and thus increase the accuracy of the parameter estimation. For this purpose we applied the Fisher formalism to the \texttt{lalsimulation} waveform \texttt{IMRPhenomHM} \cite{London:2017bcn} for the full set of 11 parameters: chirp mass $\mathcal{M}$, symmetric mass ratio $\eta$, the aligned components of the companion spins $\chi_{1,z}$ and $\chi_{2,z}$, luminosity distance $D_L$, coalescence time $t_c$, phase of coalescence $\phi_c$, inclination angle $\iota$, right ascension $\alpha$, declination $\delta$, and polarization angle $\psi$. \texttt{IMRPhenomHM} is an aligned-spin waveform model that does not include the spin components perpendicular to the orbital angular momentum of the binary, thus our Fisher analysis is four parameters short of the standard 15-parameter analyses.

Since we marginalized over the four angles for the calculation of the {\em LISA} SNRs, we randomly sampled 100 realizations of each angle for each of the 181 systems and performed the Fisher analysis on these $181\times100$ parameter sets. The resulting error bounds are shown in Fig. \ref{fig:measureability}, where we show the errors on right ascension and declination combined in the 90\%-credible sky area $\Omega_{90}$ and omitted the error for phase of coalescence.

The {\em LISA} parameter estimation has been explored with the Fisher formalism in \cite{Sesana:2016ljz} for signals with $\rho_{\it LISA}>8$. The study reports the following bounds for the majority of its 1000 simulated events: $\Delta \mathcal{M}/ \mathcal{M} \in [10^{-7},4\times10^{-6}]$ peaked at $\sim10^{-6}$, $\Delta \eta/ \eta \in [6\times10^{-4},3\times10^{-2}]$ peaked at $\sim8\times10^{-3}$, $\Delta t_c \in [10^{-1},7\times10^{1}]$ peaked at $\sim3\times10^{0}$, and $\Omega_{90} \in [2\times10^{-2},4\times10^{0}]$ peaked at $\sim\times10^{-1}$.

Comparing our error bounds to these estimates we can clearly see that a network of CE and ET observatories will outperform {\em LISA} for the estimation of most parameters: our 90\% values are either well below ($t_c$) or of the order of ($\Omega_{90}$) the lower bound of the reported ranges. The exceptions are the chirp mass and symmetric mass ratio which benefit from the long, many-cycle signals in the {\em LISA} band: the fractional $\mathcal{M}$ errors in the {\em LISA} band are better or the same compared to the 3G bounds and if we scale our absolute errors in $\eta$ to relative errors---i.e. multiplication with factors between 4 to 17 in the case of our binaries with $\eta\in[0.06,0.25]$---we obtain roughly the same ranges (the 3G network performs better on the lower end). Thus, LISA can only improve the chirp mass measurement, without adding significant information to the estimation of $\eta$.

There are two caveats to this comparison that favor the 3G results even more: The cited study performed the Fisher analysis only for six parameters which positively biases their results in comparison to our analysis over 11 parameters. Further, their reported errors come from a louder population with $\rho_{\it LISA}>8$, whereas most of our signals have SNRs lower than that. The events considered in their study would result in even louder 3G events, as seen in Fig. \ref{fig:visibility}, and thus better error estimates.

In conclusion, our findings show that a 3G network allows to estimate the parameters of \hbbh{s} with such high fidelity that we can assume most parameters to be known and focus the archival searches on $\mathcal{M}$ only.

\section{Mapping the {\em LISA} data analysis problem to the audio band}
\label{sec:mapping}
The LSC Algorithm Library (LALSuite) \cite{lalsuite} has many data analysis tools such as compact binary waveform models, template placement algorithms, filtering routines, etc., that are extremely useful, sometimes critical, in evaluating data analysis problems such as the ones explored in this paper. LALSuite was developed primarily for the analysis of data from LIGO and Virgo interferometers that operate in the audio frequency band from~1 Hz to 10 kHz. Unfortunately, some of the algorithms do not readily work at frequencies below 1 Hz and the effort required to rewrite the algorithms for the {\em LISA} band, $\sim 100\,{\rm \mu Hz}$ to $100 \,{\rm mHz}$, would be formidable.  Luckily, it is possible to scale the {\em LISA} problem into the audio band owing to the fact{} that the fundamental quantity of interest, namely the strain measured by the gravitational-wave detectors which represents the change in proper length between `free' test masses in response to a passing gravitational wave, has no physical dimension.

To illustrate the required scaling, let us consider gravitational waves emitted by an inspiraling binary system composed of a pair of black holes, but the argument would work no matter what type of source we consider. Furthermore, for the sake of clarity, we will consider the lowest order post-Newtonian (PN) waveform~\cite{Blanchet:1995ez, Buonanno:2009zt} (often referred to as the `Newtonian' waveform) from a binary system composed of non-spinning black holes. However, the arguments follow through irrespective of the PN order. At the Newtonian order, the strain response of an interferometer  to gravitational waves from a binary system composed of black holes of masses $(m_1, m_2)$ at a luminosity distance $D_L$ 
is given by 
\begin{equation}
\label{eq:strain}
    h(t) = \frac{4G\eta M}{c^2D_L} \left ( \frac{\pi G M f(t)}{c^3}  \right )^{2/3} \cos \left ( \pi \int_{t_0}^t f(t)\,\mathrm{d}t \right ) 
\end{equation}
where $M=(m_1+m_2)$ is the total mass of the system, $\eta=m_1m_2/M^2$ is the symmetric mass ratio, and $f(t)$ is the instantaneous gravitational-wave frequency:
\begin{equation}
\label{eq:frequency}
    f(t) = f_0 \left ( 1 - \frac{t-t_0}{\tau} \right )^{-3/8}.
\end{equation} 
Here $t_0$ is a fiducial time when the frequency of the gravitational wave is $f_0$ and $\tau,$ called the \emph{chirp time}, is the duration of the signal from a time when its frequency was $f_0$ until the frequency (in this approximation) diverges\footnote{In reality, the merger occurs when the horizons of the two black holes merge which happens at a finite frequency but this is not relevant to our discussion.} and the two black holes merge: 
\begin{equation}
\label{eq:chirp time}
    \tau = \frac{5}{256\,\eta}\frac{GM}{c^3} \left ( \frac{c^3}{\pi\, G M f_0} \right )^{8/3}.
\end{equation}
The chirp time of a binary of total mass $100\,M_\odot$, equal component masses (so that $\eta=1/4$), and starting frequency of 12 mHz would be 5 years.  Chirp time is a sharp function of the total mass as well as the starting frequency. A signal starting from a frequency that is a factor 2 (10) smaller would last a factor $\sim 6.3$ (respectively, 464) longer.  

From Eq.~\eqref{eq:chirp time}, we can compute the duration $\Delta t$ spent by the binary starting at frequency $f_1$ at time $t_1$ and reaching frequency $f_2$ at time $t_2$:
\begin{equation}
\label{eq:signal duration}
    \Delta t \equiv t_2 - t_1 = \frac{5}{256\,\eta}\frac{GM}{c^3} 
    \left ( \frac{c^3}{\pi\, G M f_1} \right )^{8/3} 
    \left [ 1 - \left ( \frac{f_1}{f_2} \right)^{8/3} \right ].
\end{equation}
If $f_2\gg f_1$, the second term in the equation above will be negligible, and $\Delta t$ will essentially be the same as the chirp time starting at frequency $f_1$:
\begin{equation}
\label{eq:signal duration approximate}
    \Delta t \simeq \frac{5}{256\,\eta}\frac{GM}{c^3} \left ( \frac{c^3}{\pi\, G M f_1} \right )^{8/3}.
\end{equation}
We choose the starting frequency $f_1$ for the stellar mass binary black holes in the mass range observed by {\em LISA}, such that the signal lasts for a fixed duration in the {\em LISA} band. For a given $\Delta t$, the starting frequency $f_1$ depends on the total mass and mass ratio of the binary:
\begin{equation}
\label{eq:fstart}
    f_1(M,\eta) = \frac{c^3}{\pi\, G M} \left [ \frac{5}{256\,\eta}\frac{GM}{c^3 \Delta t} \right ]^{3/8}.
\end{equation}
\begin{figure*}
    \centering
    \includegraphics[width=\textwidth]{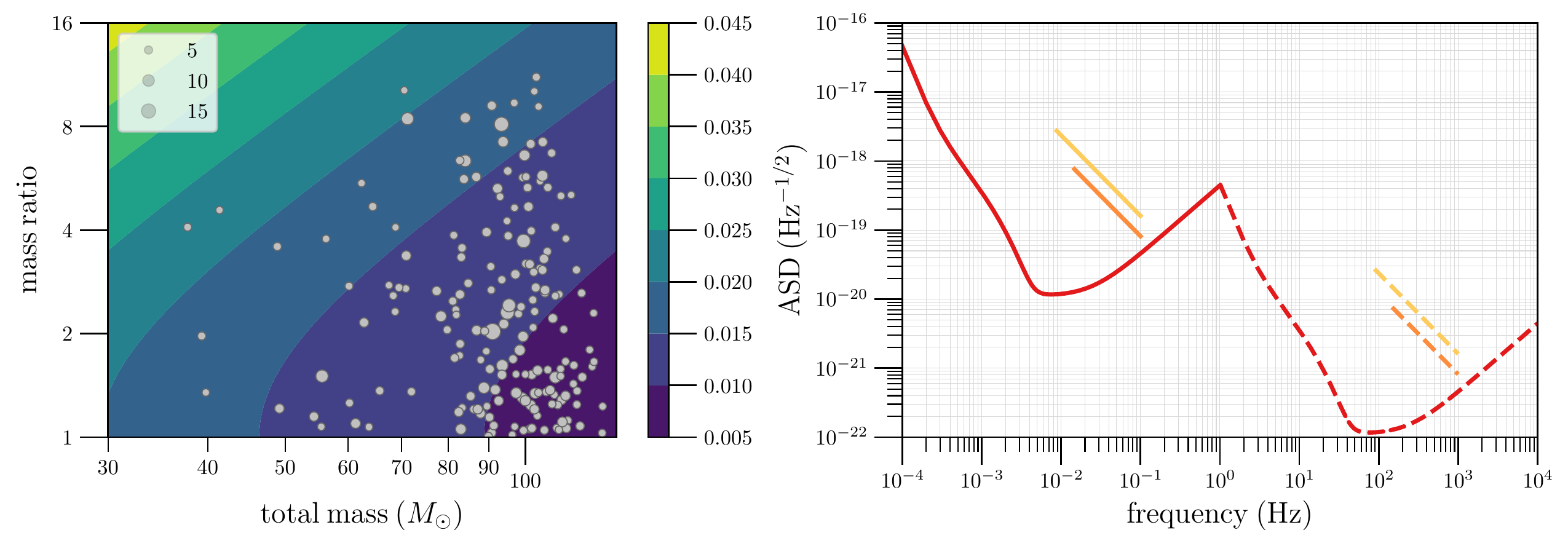}
    \caption{{\bf Left:} Contour plot of the starting frequency for stellar mass binary black holes that last 5 years in the {\em LISA} band. The starting frequency scales with the mass ratio of each system and inversely with the total mass, as indicated by the colored contours, where the values of starting frequency are shown in mHz. The grey dots mark the positions of the 181 \hbbh{s} for which we performed Fisher analysis in a 3G network to obtain parameter error estimates. The size of the dots represents the corresponding {\em LISA} SNR, $\rho_{\rm {\it LISA}}$.\\
    {\bf Right:} {\em LISA}'s amplitude noise spectral density (ASD) $\sqrt{S_h(f)}$ is plotted before (red solid line) and after (red dashed line) applying frequency scaling with $\alpha = 10^4$. Also shown are the amplitude spectra $\sqrt{f}|\tilde h(f)|$ for a GW150914-like (in orange) and GW190521-like (yellow) systems at a distance of 500 Mpc with signal-to-noise ratio of 4.8 and 16, respectively. The integration time is assumed to be 5 years in the {\em LISA} band which translates to $1.54\times 10^5$ s in the audio band.}
    \label{fig:fstart}
\end{figure*}
The left half of  Fig.\ \ref{fig:fstart} shows the starting frequency $f_1$ as a function of total mass and mass ratio $q=m_1/m_2$ for $m_1>m_2$; $f_1$ is greatest for systems with small total mass but large mass ratio and smallest for systems with large total mass, but small mass ratio. Over the total mass range of [30, 130] $M_{\odot}$ the lower frequency cutoff is never smaller than 5 mHz and can be as large as 45 mHz at the lower end of the mass range and upper range of the mass ratio. The upper frequency cutoff for integration is chosen to be $f_2=1\,\rm Hz.$

Our goal is to scale up the frequency from {\em LISA}'s observing band to the audio band of ground-based detectors. Scaling up frequencies by a factor of $\alpha=10^4$ would bring the lowest starting frequencies up to 50 Hz and the largest starting frequencies up to 450 Hz. This is the scaling we will use in this paper. An integration time of $\Delta t_{\rm {\it LISA}}=5\,\rm yr$ in the {\em LISA} band would correspond to an integration time of $\Delta t_{\rm audio}=\Delta t_{\rm {\it LISA}}/\alpha \simeq 1.58\times 10^4\,\rm s.$

We can see from Eq.\,(\ref{eq:strain}) that gravitational-wave strain would remain unchanged if we simultaneously scale up the frequency by factor of $\alpha$ and scale down the chirp time, the total mass, and the luminosity distance by the same factor. Therefore, the signal would now last for a much shorter period of $\Delta t /\alpha$ with exactly the same amplitude as before but at a higher frequency. The SNR of the scaled up signal, but with a scaled up {\em LISA} noise spectral density $S_h^{\rm {\it LISA}}(f)$, will also be the same as before. To see this, recall that the expectation value of the SNR of a signal is given by: 

\begin{equation}
    \label{eq:SNR}
    \rho^2(f_1,f_2) = 4 \Re \int_{f_1}^{f_2} \frac{|\tilde h(f)|^2}{S_h^{\rm {\it LISA}}(f)}\,\mathrm{d}f,
\end{equation}
where $\tilde h(f)\equiv \int_{-\infty}^\infty h(t)\,\exp(2\pi i f t)\,\mbox{d}t$ is the Fourier transform of the gravitational-wave strain. Changing the variable $f\rightarrow \nu = \alpha f$ would scale down the Fourier mode strain by a factor of $\alpha$, $\tilde h(f) \rightarrow \tilde h(\alpha f)/\alpha$, and similarly the {\em LISA} noise PSD $S_h^{\rm {\it LISA}}(f) \rightarrow S_h^{\rm {\it LISA}}(\alpha f) / \alpha \equiv  S_h^{\rm audio}(\alpha f) / \alpha$. Thus the SNR remains unchanged:
\begin{equation}
    \label{eq:SNR-scaled}
    \rho^2(\nu_1,\nu_2) = 4 \Re \int_{\nu_1}^{\nu_2} \frac{|\tilde h(\alpha f)|^2 / \alpha^2}{ S_h^{\rm audio}(\alpha f)/\alpha}\, \alpha \mathrm{d}f.
\end{equation}
The scaled version of the {\em LISA} noise PSD $S_h^{\rm audio}(\nu)/\alpha$ is shown in the right hand panel of Fig. \ref{fig:fstart}.     

In summary, the required scaling transformation are to: 
\begin{enumerate}
    \item Scale up the frequency: $f\rightarrow \alpha f.$
    \item Scale down the time duration, total mass and distance:
    $\tau \rightarrow \tau/\alpha$, $M\rightarrow M/\alpha,$ and $D_L \rightarrow D_L/\alpha$.
    \item Transform the {\em LISA} power-spectral density into the audio band, i.e. $S_h^{\rm audio}(f) = \frac{1}{\alpha} S^{\rm {\it LISA}}_h(f/\alpha)$.
\end{enumerate}

\section{Template banks for archival searches}
\label{sec:archival searches}

In this section we present the number of templates required to detect \hbbh\, in the {\em LISA} band by matched filtering. We use accuracies on the masses obtained from parameter estimation in 3G detectors. We use two independent methods to calculate template bank numbers. The first method assumes the metric in order to place templates, which would provide a minimum on the number of templates required. As a check on this method, we also calculate the number of templates using a stochastic placement algorithm, which overestimates the required number of templates.
\subsection{Metric method}
\subsubsection{Metric on the signal manifold}
\label{sec:num templates}
The number of templates required for a search can be found using the geometric formulation of signal analysis \cite{Balasubramanian:1995bm, Owen:1995tm, Owen:1998dk}. The scalar product \eqref{eq:scalar_product} can be used to define waveforms or signals of {\em unit norm}. A signal is said to be of unit norm if its scalar product with itself is unity and will be denoted by $\hat a:$ $\langle\hat a,\hat a\rangle =1.$ 

In the geometric formalism, the overlap $\cal O$ or {\em match} between two `nearby' normalized signals $\hat h(\boldsymbol{\lambda})$ and $\hat h(\boldsymbol{\lambda}+\Delta\boldsymbol{\lambda})$ with slightly different parameters $\boldsymbol{\lambda}$ and $\boldsymbol{\lambda}+\Delta\boldsymbol{\lambda}$ is given by:
\begin{equation}
    {\cal O}(\boldsymbol{\lambda},\Delta\boldsymbol{\lambda}) 
    \equiv  \langle \hat h(\boldsymbol{\lambda}), \hat h(\boldsymbol{\lambda}+\Delta \boldsymbol{\lambda}) \rangle \approx 1 - g_{\alpha\beta}\, 
    {\rm d}\lambda^\alpha\,{\rm d}\lambda^\beta,
\label{eq:match}
\end{equation}
where $g_{\alpha\beta}$ is the metric on the signal manifold \cite{Balasubramanian:1995bm,Owen:1995tm}:
\begin{equation}
    g_{\alpha\beta} \equiv -\frac{1}{2} \left . \frac{\partial^2 {\cal O}(\boldsymbol{\lambda},\Delta\boldsymbol{\lambda}) 
    }{\partial\lambda^\alpha\partial\lambda^\beta} \right |_{\Delta\boldsymbol{\lambda}=0} = \langle  \hat h_\alpha, \hat h_\beta \rangle, \quad \hat h_\alpha \equiv \frac{\partial \hat h}{\partial \lambda^\alpha}.
\label{eq:metric}
\end{equation}
When signals are nearby, in the sense that their overlap is close to unity, Eq.\,(\ref{eq:match}) is a good approximation for the overlap and the quantity ${\rm d}\ell^2=g_{\alpha\beta}\, {\rm d} \lambda^\alpha \, {\rm d}\lambda^\beta$---the {\em proper distance} between them---obeys ${\rm d}\ell \ll 1.$ Thus, two normalized signals at a proper distance of $d\ell$ from each other have an overlap of $1-{\rm d}\ell^2.$ 

\subsubsection{Minimal match and lower limit on the number of search templates}

To filter signals out of data we must choose a bank of templates in the parameter space of interest such that any signal buried in the data within this parameter space has its overlap larger than a certain value called the {\em  minimal match} $M$ with at least one template in the bank. Of course, this requirement can be met by populating the parameter space with a dense set of templates but a higher density of templates would demand a greater computational cost. Thus, the density of templates must be chosen so that it is as sparse as possible while assuring minimal match with every signal in the parameter space of interest. 

If $\boldsymbol{\lambda_k},$ $k=1,\ldots N,$ denotes the parameters of the templates in the bank then for a signal with arbitrary parameters $\boldsymbol{\lambda}$ the template bank must satisfy the following condition:
\begin{equation}
    \max_k \langle \hat h(\boldsymbol{\lambda}), \hat h(\boldsymbol{\lambda}_k)  \rangle \ge M,
\end{equation}
the equality in the above equation giving the optimal number of templates. So we must choose the proper distance such that
\begin{equation}
    {\rm d}\ell^2 = g_{\alpha\beta}\,{\rm d}\lambda^\alpha\,{\rm d}\lambda^\beta = (1-M).
    \label{eq:proper distance}
\end{equation}
Given that the optimum matched filter signal-to-noise ratio that one can hope to achieve for a signal is $\rho_{\rm opt}^2 = \langle h,h\rangle,$  the above condition assures that the fractional drop in the signal-to-noise ratio between an arbitrary signal and the closest template in the bank is no more than $\epsilon \equiv 1-M$ called the {\em maximum mismatch,} i.e. $\rho\ge \epsilon \rho_{\rm opt}.$ 

Given the minimal match $M$ how many templates are needed to cover the parameters with the smallest number of templates? This is the problem of template placement and to some extent the answer depends on what type of lattice is used to place the templates on the signal manifold. We will discuss a specific template placement algorithm used in this work in the next Section. We can get a lower limit on the number of templates needed by computing the total proper volume of the signal space divided by the fraction of volume covered by each template. Assuming that each template covers a proper volume ${\rm d}V={\rm d}\ell^N=(1-M)^{N/2},$ where $N$ is the dimension of the parameter space, the smallest number of templates $N_T$ needed is:
\begin{equation}
    N_T = \frac{V}{{\rm d}V} = \frac{1}{(1-M)^{N/2}} \int^{\boldsymbol{\lambda}_{\rm max}}_{\boldsymbol{\lambda}_{\rm min}} \sqrt{g}\,{\rm d}^N\lambda,
    \label{eq:number of templates}
\end{equation}
where $g=\det\left | g_{\alpha\beta}\right |.$ The above estimate assumes that the templates are placed on a square lattice. The number of templates can be smaller with a more efficient lattice, e.g. a hexagonal lattice in two-dimensions, but this is unimportant for our considerations.

\subsubsection{Newtonian approximation to number of templates}
The orbital velocity of a stellar mass binary black hole of total mass $M$ and gravitational-wave frequency $f_{\rm GW}$ is very small in the {\em LISA} frequency band compared to the speed of light (for clarity we include factors of $G$ and $c$ in the equation below): 
\begin{equation}
    \frac{v}{c} \equiv \left (\frac{\pi G M f}{c^3} \right )^{1/3} \simeq 0.025\,\left ( \frac{M}{10^2M_\odot} \right )^{1/3} \left ( \frac{f_{\rm GW}}{10\, \rm mHz}\right )^{1/3}.
\end{equation}
For such a non-relativistic orbit the PN expansion parameter defined by $x\equiv v^2/c^2 \simeq 6.2 \times 10^{-4},$ meaning
stellar-mass binary black holes in the {\em LISA} frequency band are in the adiabatic regime. The dynamics of such a system is essentially described by the Newtonian approximation (or 0 PN) given in Eqs.\,(\ref{eq:frequency}) and (\ref{eq:chirp time}). In the stationary phase approximation the Fourier transform of the waveform in Eq. (\ref{eq:strain}) is given by \cite{Sathyaprakash:1991mt}:
\begin{equation}
    h(f) = \frac{A}{D_L} f^{-7/6} \exp\left [ 2\pi i f t_C - i\phi_C + i\psi(f) \right ],
\end{equation}
where $t_C$ is a fiducial time giving the time of arrival of the signal at the detector (the epoch at which the GW frequency reaches a pre-defined value), $\phi_C$ is a constant phase of the signal, and $\psi(f)$ is the PN approximation to the signal's phase evolution given at the leading ``Newtonian order'' by: 
\begin{equation}
    \psi(f) = \frac{3}{128(\pi {\cal M}f)^{5/3}}.
\end{equation}
At the leading order the phase depends only on the chirp mass $\cal M$ and not the mass ratio. It is useful to define a new parameter $\xi\equiv (3/128)(\pi\,{\cal M})^{-5/3},$ so that the phase is linear in this new parameter: $\psi(f)=\xi f^{-5/3}.$ The parameter space of the signal consists of $\boldsymbol{\lambda}=(t_C, \phi_C, \xi).$ It turns out that the overlap $\cal O$ can be analytically maximized over the phase $\phi_C$ by using two quadrature filters, leaving just two parameters. Because of the analytic maximization over phase, the expression for the metric in the two-dimensional space of $\boldsymbol{\lambda}=(t_C,\xi)$ takes the form \cite{Owen:1995tm, Owen:1998dk}:
\begin{equation}
   g_{\alpha\beta} = \frac{1}{2}\left ( {\cal J}[\psi_\alpha\psi_\beta] - {\cal J}[\psi_\alpha]{\cal J}[\psi_\beta] \right ), 
   \label{eq:metricJ}
\end{equation}
where $\psi_\alpha \equiv \partial \psi(f)/\partial\lambda^\alpha,$ and ${\cal J}$ is a functional of its argument defined for any function $a(f)$ by 
\begin{equation}
    {\cal J}[a(f)] \equiv 
    \frac{1}{\rho^2} \langle f^{-7/3}, a(f) \rangle
\end{equation}
As is well known, maximizing the overlap $\cal O$ over the parameter $t_C$ is easily accomplished in the Fourier domain \cite{Schutz:1989cu} and one needs a discrete lattice of templates only for the remaining one parameter $\xi.$ The metric in the $\xi$-dimension is quite simply:
$G_{22} = g_{22} - g_{21}^2/g_{11}.$ Substituting for the various elements of the metric $g_{\alpha\beta}$ and simplifying one finds:
\begin{equation}
    G_{22} = \frac{1}{2}\left [ J_{17} - J_{12}^2 - \frac{(J_9 - J_4 J_{12})^2}{J_1-J_4^2} \right ]
\end{equation}
where for any $k,$ the moment $J_k$ is given by $J_k \equiv \langle f^{-k/3}, 1 \rangle.$ In this notation the SNR is $\rho^2=J_7.$ $G_{22}$ is demonstrably constant (in any case all one-dimensional spaces are flat) and $\xi$ is already a Cartesian coordinate. The spacing between templates is constant in $\xi$ and from Eq.\,(\ref{eq:proper distance}) we have
\begin{equation}
    {\rm d}\ell^2 = G_{22}\, {\rm d}\xi^2 = 1 - M \quad \Rightarrow \quad {\rm d}\xi = \sqrt{\frac{1-MM}{G_{22}}}
\end{equation}
Finally, the number of templates can be found using Eq.\,(\ref{eq:number of templates}):
\begin{equation}
    N_T = \sqrt{\frac{G_{22}}{(1-M)}} (\xi_{\rm max} - \xi_{\rm min}).
\end{equation}
In the next section we will compute the number of templates found using a template placement algorithm and compare it with the one found in this section.

\begin{figure*}
    \centering
    \includegraphics[width=0.49\textwidth]{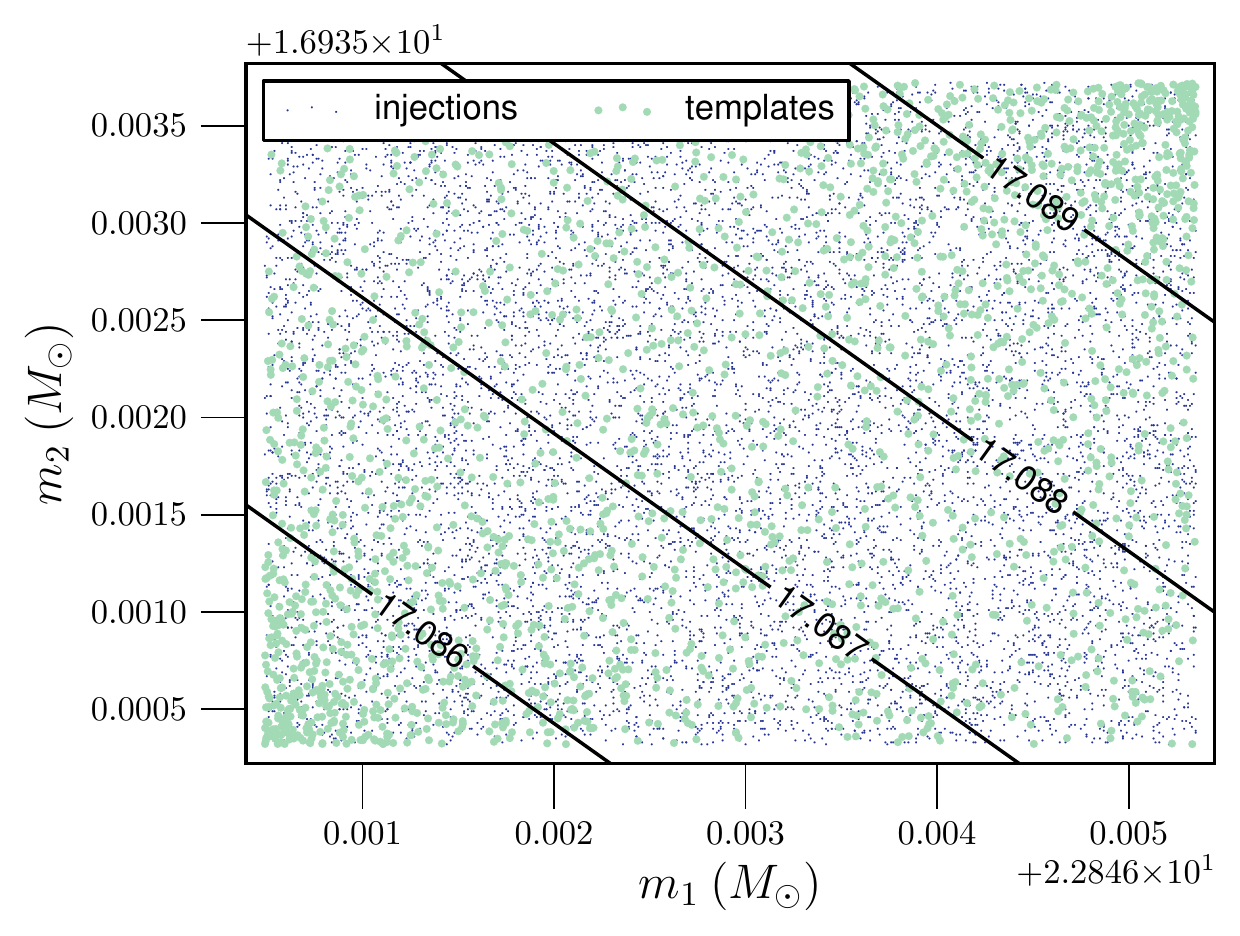}
    \includegraphics[width=0.49\textwidth]{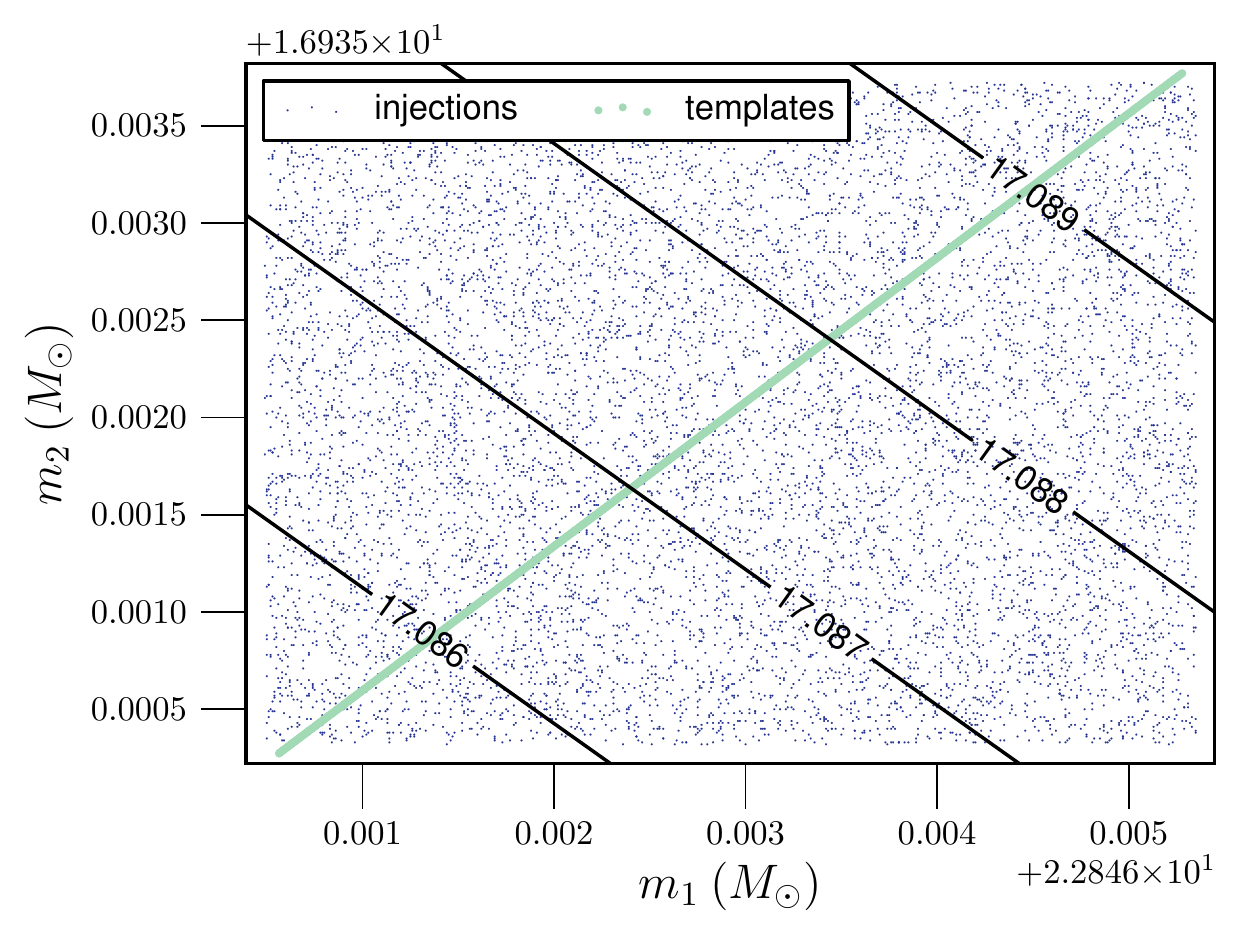}
    \caption{{\bf Left:} For a \hbbh\ source with component masses constrained by 3G to $m_1 = 22.84892 $ and $m_2 = 16.93702$ $M_{\odot}$ the template (green) and injection (blue) component masses are shown for a 3.5 PN template bank. chirp mass contours are indicated by the black lines. {\bf Right:} For the same source, the injection component masses are shown with the one-dimensional Newtonian template component masses. }
    \label{fig:tmplt_inj_m1_m2}
\end{figure*}

\subsection{Stochastic template placement algorithm}

When the metric on the waveform manifold is not known exactly and cannot be approximated, we use a brute force approach~\cite{Harry:2009ea, ajith2014effectual} to construct the template bank. In this method, the template bank is built by proposing templates in the desired parameter space until sufficient coverage is reached. For each proposed template, $\boldsymbol{\lambda_{prop}}$, the fitting factor (which is the maximum match of the proposed template with the templates in the bank) is calculated. If the fitting factor is greater than the minimal match required, we reject the proposed template and continue with a new proposal; if it is less than the required minimal match, we add the proposed template to our bank. 

This method can get computationally expensive, but there are tricks we can employ to be able to use it in practice.
One such trick is to define the ``neighborhood" of the proposed templates. It can be defined in terms of a parameter chosen by the user, we used the chirp time for our purposes. When used, the fitting factor is calculated only for the templates in the neighborhood (set in form of units of the neighborhood parameter by the user), 
$\boldsymbol{\lambda_{i}}$, where $i: 1 - N$ for N templates near the proposal. 

Another technique we use is to iteratively lower the frequency step in the calculation of the match~\cite{Capano_2016}. The value of frequency spacing used in the match integral (Eq.~\eqref{eq:scalar_product}) is usually chosen to be $1/L$, where $L$ is the closest power-of-2 greater than the length of the waveforms. This is required to measure the overlap between two waveforms in a time window of $L \,\rm s$. But for the bank generation, we are interested in the maximum overlap between waveforms, which occurs near the time point corresponding to 0 displacement for two waveforms aligned at their peak amplitude. Therefore, we can increase the value of frequency spacing, $df$, which greatly reduces the cost of calculating the match. We can check if our chosen value for $df$ is good enough by iteratively calculating matches by reducing $df$ by half. If the last two overlaps agree to within 1\%, we use that value or if $(1 - \rm(match))$ is large, we continue to the next template. For our banks, we first calculate the matches with $df=2.0$. If the mismatch is large, i.e.  
\begin{equation}
    (1 - \mathrm{match}) > 0.05 + (1 - M),
\end{equation}
where $\mathrm{M} = 0.98$ is the minimum required match for template placement, then we move on to the next neighboring template $\boldsymbol{\lambda_{i+1}}$. Otherwise, we decrease the value of $df \rightarrow df/2$ and calculate the match again. We continue iteratively decreasing $df$ until the last two matches converge, 

\begin{equation}
    \left| \mathrm{match}_{df} - \mathrm{match}_{df/2} \right| < 0.001*\mathrm{match}_{df}.
\end{equation}
Once convergence is reached, if the $\mathrm{match} < M$ we add the proposal $\boldsymbol{\lambda_{prop}}$ to the template bank. 

\subsection{Number of templates}

\begin{figure}
    \centering
    \includegraphics[width=0.5\textwidth]{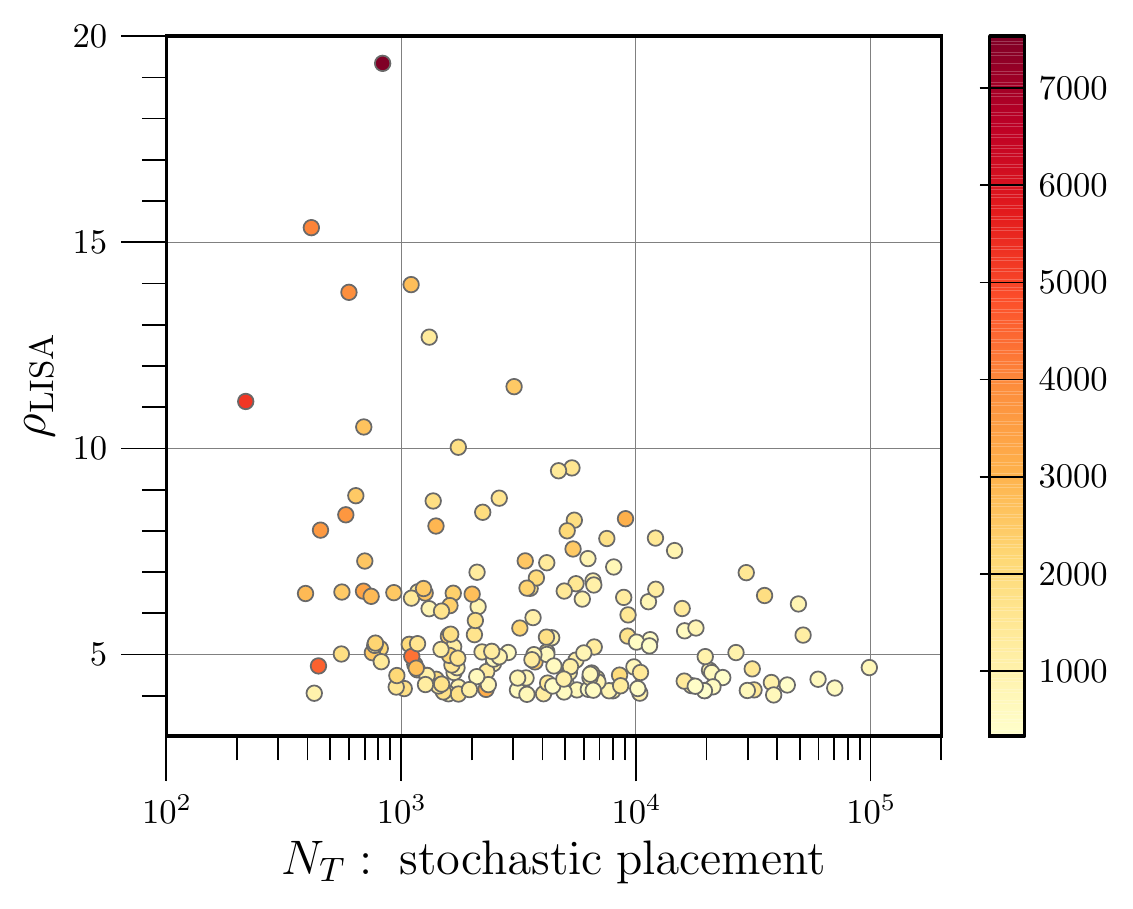}
    \caption{The number of templates from the stochastic placement algorithm is plotted against the {\em LISA} SNR, $\rho_{{\it LISA}}$ for each source. The color bar indicates the CE SNR, $\rho_{CE}$. }
    \label{fig:num_tmplts_vs_snr}
\end{figure}

\begin{figure}
    \centering
    \includegraphics[width=0.5\textwidth]{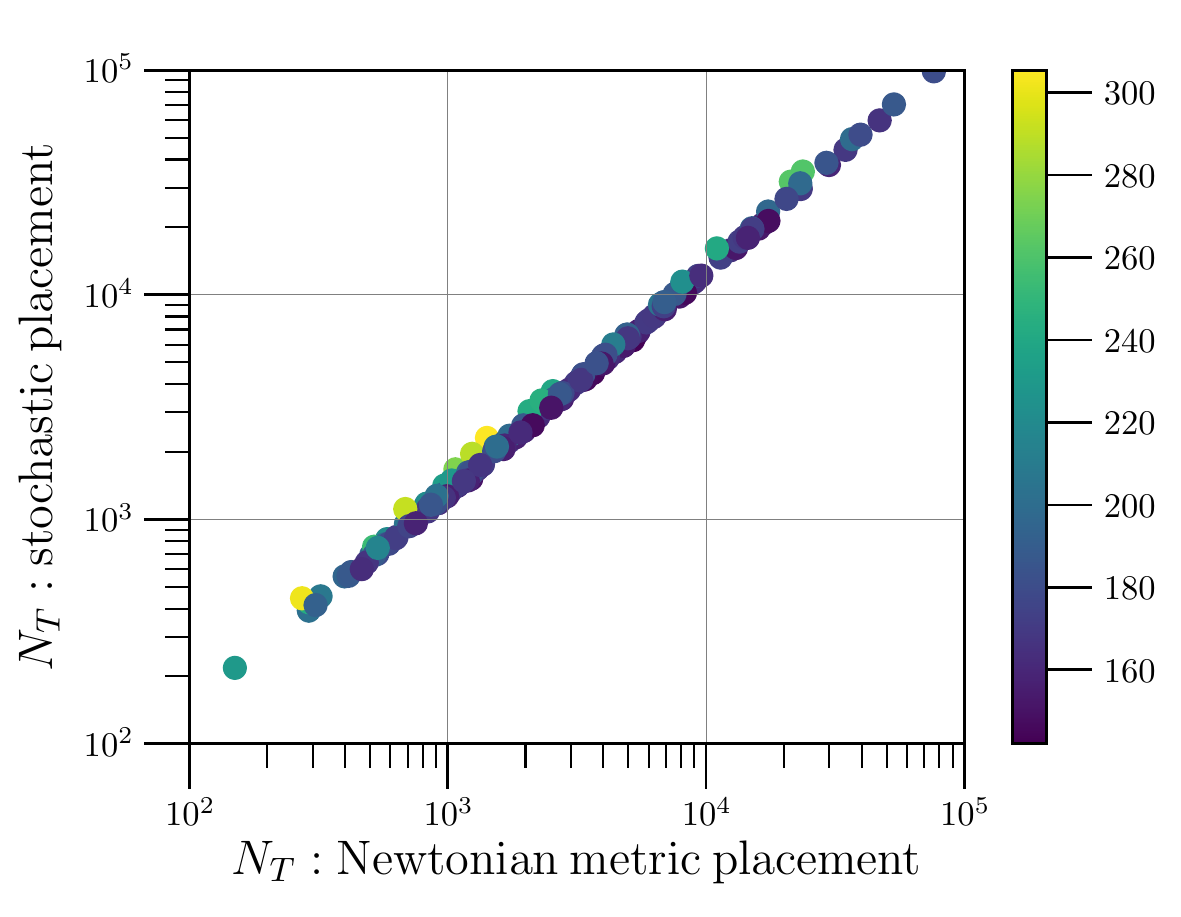}
    \caption{For 181 \hbbh\ sources we plot the number of templates required for the minimal match condition, $M = 0.98$ calculated using the metric approach for one-dimensional Newtonian template banks (horizontal axis) and the number of 3.5 PN templates found using the stochastic placement algorithm (vertical axis). The colorbar indicates the lower frequency cutoff, after scaling to the audio band. For a linear fit, $N_{T, \mathrm{stochastic}} = m \times N_{T, \mathrm{metric}}$, we find $m = 1.30$ }
    \label{fig:bank_sim_vs_newtonian_flow}
\end{figure}

Using the stochastic placement algorithm, we calculate a template bank for each of the 181 \hbbh\ sources, with component mass ranges constrained by 3G capabilities. These template banks use 3.5 PN TaylorF2~\cite{Sathyaprakash:1991mt, 1998PhRvD..57.5287P, Mikoczi:2005dn, Bohe:2013cla, Bohe:2015ana, Arun:2008kb, Mishra:2016whh} waveforms. The lower frequency cut-off for the signals ranges from 14.2--30.5 mHz with an upper frequency cut-off of 150 mHz. Scaling this to the audio band, we produce template banks with frequencies in the range of 142--1500 Hz. We use a minimal match of $M=0.98$ for the template placement algorithm. 

The left panel of Fig. \ref{fig:tmplt_inj_m1_m2} shows an example template bank in $(m_1, m_2)$ space for a source of $m_1 = 22.85 M_{\odot}$, $m_2 = 16.94 M_{\odot}$, and $f_{low} = 305.44$ Hz, after scaling to the audio band. Fig. \ref{fig:num_tmplts_vs_snr} shows the relationship between the source signal-to-noise ratio and the number of templates in the bank. The plot shows that the sources which require the largest number of templates are those for which both the {\em LISA} and CE signal-to-noise ratio is small. 

The cumulative distribution of stochastic placement template bank sizes for all 181 sources is shown in the top left panel of Fig. \ref{fig:bank-eff-results}. The 50th percentile and 90th percentile bank sizes are $3.4 \times 10^3 $ and $1.95 \times 10^4$ templates, respectively. The mean bank size, $\overline{N_T} = 8.02 \times 10^3$ with a standard deviation of $9.70 \times 10^2$. The smallest bank has only 218 templates while the largest bank has $9.88 \times 10^4$ templates. These bank sizes demonstrate a significant improvement from previous estimates of $\mathcal{O}(10^{12})$ templates. 

Using the metric placement method, we again calculate template banks for each of the same 181 \hbbh\ sources. The upper and lower frequency cut-offs are the same for these banks as for the template banks generated with the stochastic placement algorithm. We again require a minimal match, $M=0.98$. The right panel of Fig.  \ref{fig:tmplt_inj_m1_m2} shows the Newtonian template bank for the same source as shown on the left. This figure demonstrates that the Newtonian template banks are truly one-dimensional, where each template in ($m_1$, $m_2$) space is projected onto a line parameterized by the chirp mass, $\mathcal{M}$. The figure shows chirp mass contours indicated by the black lines. Using the one dimensional template bank it is clear that while component masses may not necessarily be tightly constrained by {\em LISA}, the chirp mass will be recovered very well. For this source, the two-dimensional bank had $2.3 \times 10^3$ templates and the one-dimensional bank had $1.4 \times 10^3$ templates. Therefore, the line in Fig. \ref{fig:tmplt_inj_m1_m2} is very densely packed. 

The cumulative distribution of Newtonian template bank sizes is shown in the bottom left panel of Fig. \ref{fig:bank-eff-results}. Here, the mean bank size is $6.2 \times 10^3$ templates. The 50th and 90th percentile bank sizes are $2.5 \times 10^3$ and $1.5 \times 10^4$ respectively. 

A comparison of stochastic placement and metric method template bank sizes is given in Fig. \ref{fig:bank_sim_vs_newtonian_flow}. We find that $N_{T, 2D} \approx 1.3 \times N_{T, 1D}$. The two-dimensional PN template banks are about 1.3 times the size of the one-dimensional Newtonian template banks made using the metric method. Therefore, by using Newtonian template banks we can achieve an even further decrease in the computational cost required to dig signals out of the {\em LISA} data. 

\subsection{Efficiency of the Template Bank}
\label{sec:sims}

\begin{figure*}
    \centering
    \includegraphics[width=0.49\textwidth]{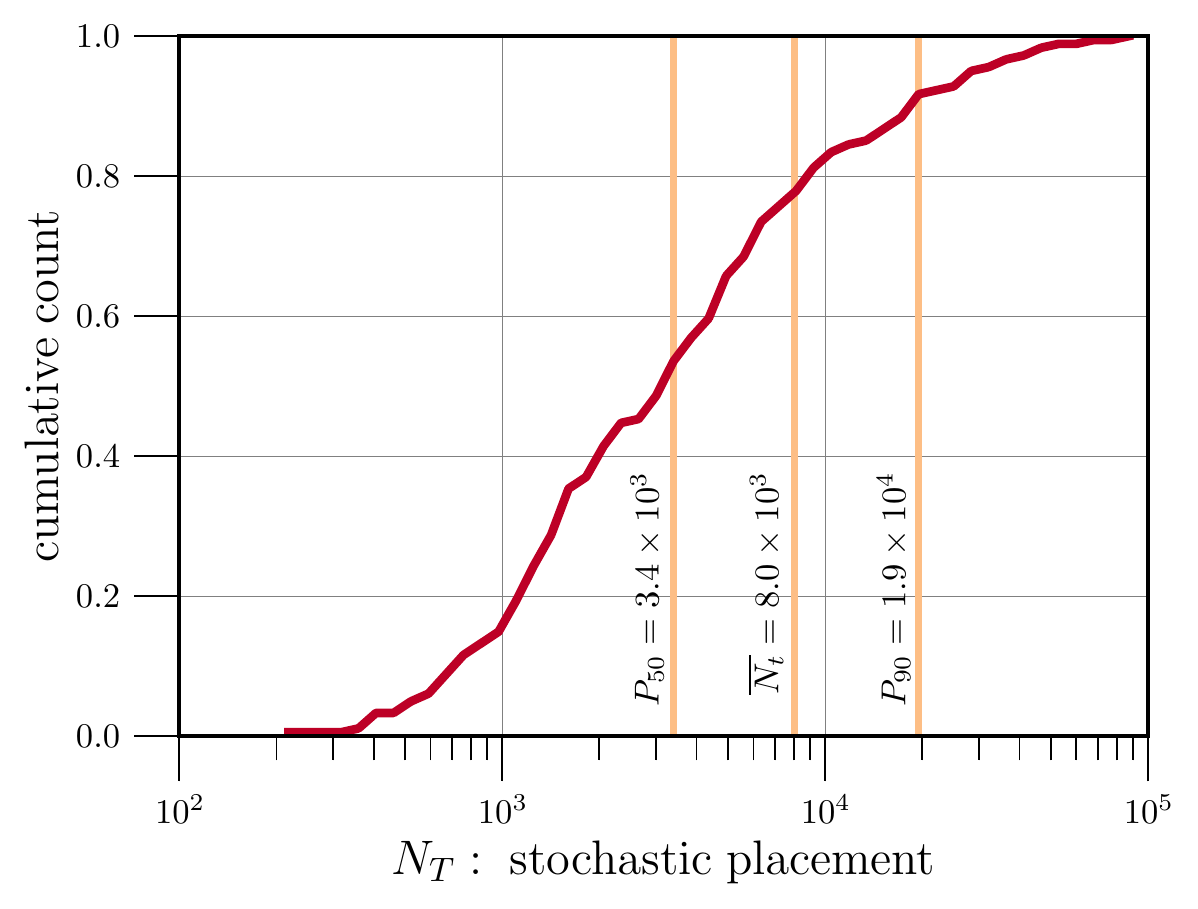}
    \includegraphics[width=0.49\textwidth]{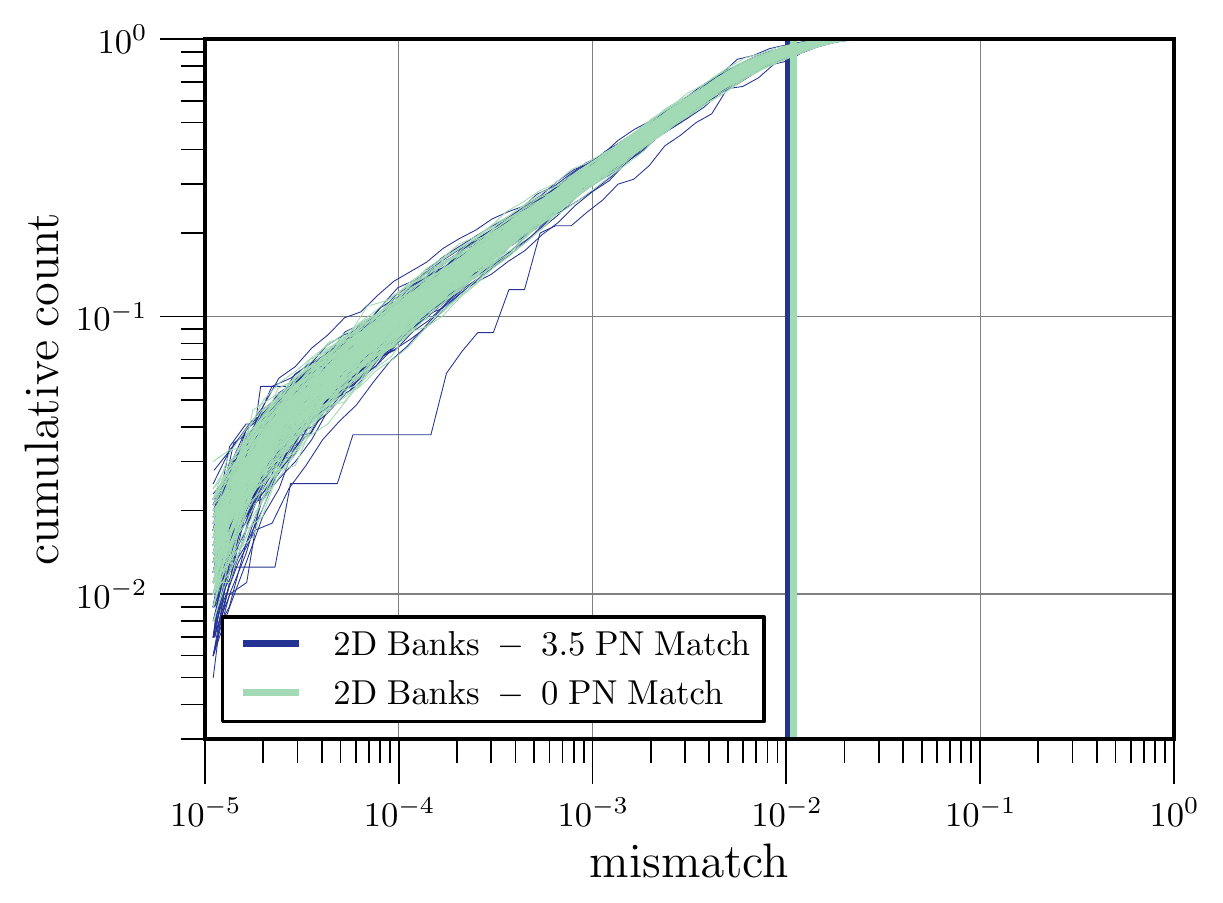}
    \includegraphics[width=0.49\textwidth]{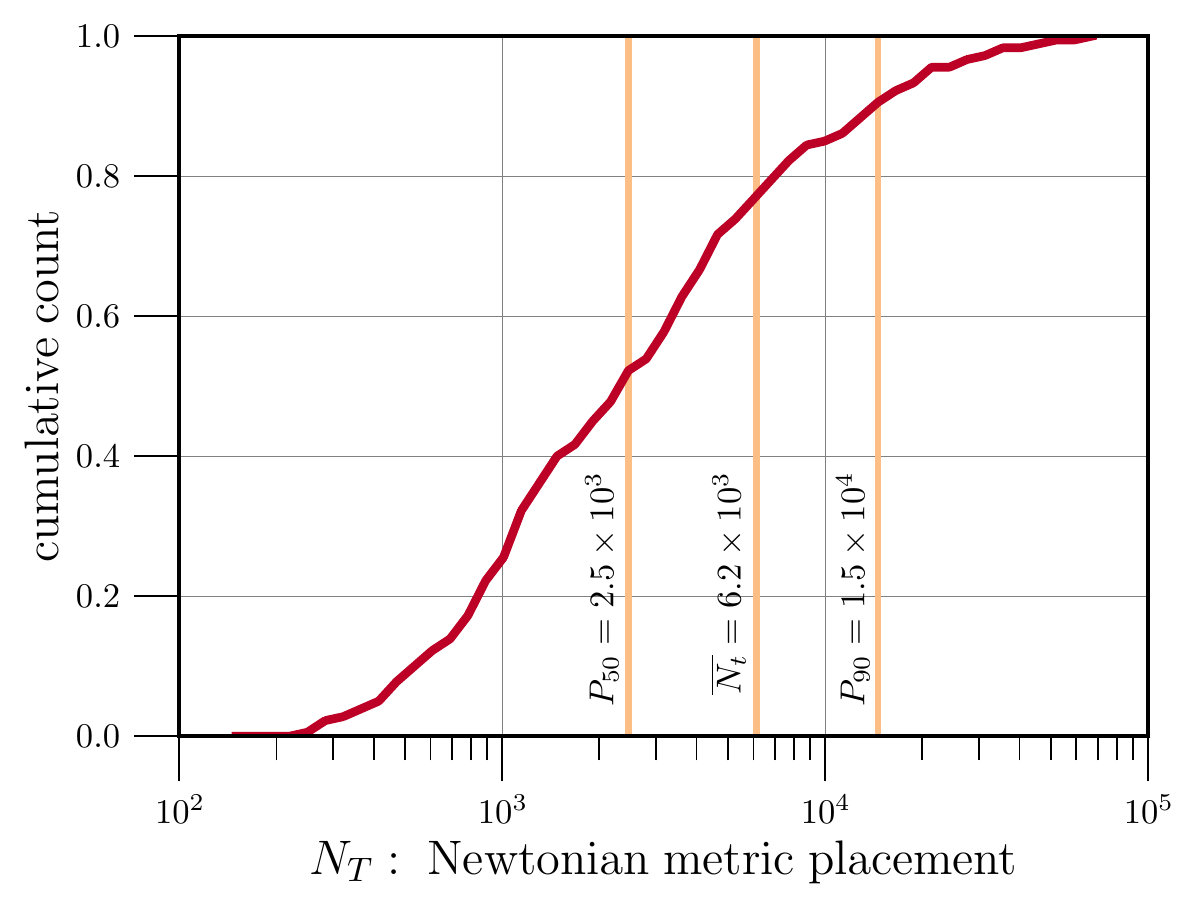}
    \includegraphics[width=0.49\textwidth]{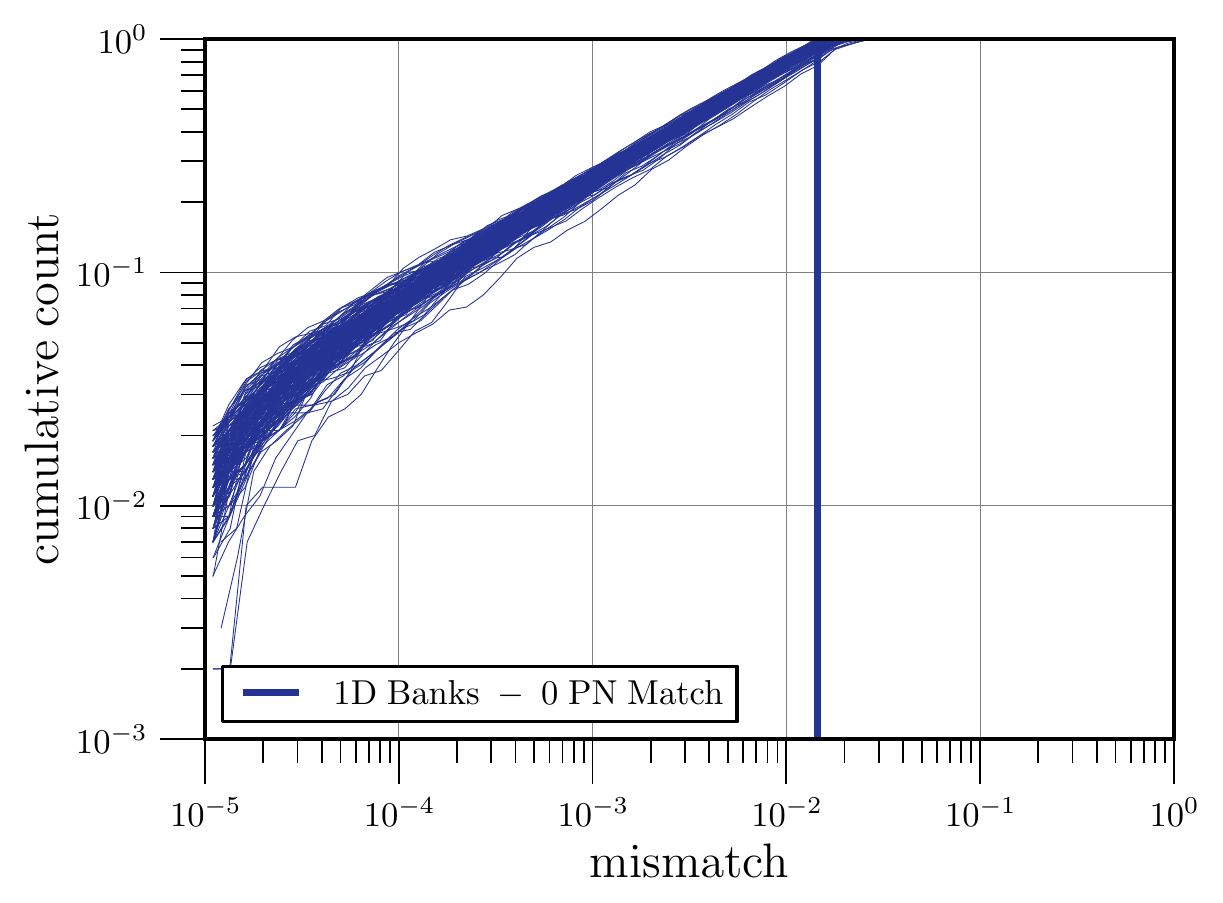}
    \caption{{\bf Top Left:} The plot shows the cumulative distribution of the 3.5 PN template bank sizes. The vertical lines indicate, from left to right, the 50th percentile, mean, and 90th percentile bank sizes. The bins are evenly spaced in log. {\bf Bottom Left:} The cumulative distribution of one-dimensional Newtonian template bank sizes. {\bf Top Right:} The plot shows the cumulative distribution of mismatches of 1000 injections with each 3.5 PN template bank. In the match calculation, we use the Newtonian approximation (green) and the 3.5 PN approximation (blue). The vertical lines indicate the mean 10th percentile match for both sets of template banks. The position of the mean 10th percentile match value for 3.5 PN approximation has been artificially shifted to the left by 0.0005 so that the two lines are distinguishable. {\bf Bottom Right:} The cumulative distribution of mismatches for one-dimensional Newtonian template banks.}
    \label{fig:bank-eff-results}
\end{figure*}

To determine the bank efficiencies, we filter 1000 3.5 PN waveform injections against each bank to find the best matching templates. For each of the two-dimensional banks, we calculate the efficiency twice -- first, using the 3.5 PN approximation in the match calculation and again using only the Newtonian approximation. A summary of all the bank efficiency calculations performed is given in Table \ref{tab:bank_eff_runs}. The top right panel of Fig. \ref{fig:bank-eff-results} shows the cumulative distribution of matches for each of the two-dimensional template banks, where green lines show matches calculated with the Newtonian template approximation and blue lines show matches calculated with the 3.5 PN template approximation. The mean $10^{\mathrm{th}}$ percentile match for both sets are indicated by the vertical lines. These results  demonstrate that using the Newtonian approximation in the match calculation is sufficient for our purposes. 

This result is further proven by the one-dimensional Newtonian template bank efficiencies, shown in the bottom right panel of Fig. \ref{fig:bank-eff-results}. As indicated by Table \ref{tab:bank_eff_runs}, for this set of bank efficiency calculations we use the Newtonian approximation both in the template waveforms and in the match calculation. Each of the three sets of bank efficiency calculations performed over the 181 \hbbh\ systems had mean $10^{\mathrm{th}}$-percentile matches $M > 0.98$, so we conclude that these template banks will be effective in recovering \hbbh\ signals from {\em LISA} data. 

\begin{table}
\begin{tabular}{c c c c}
  $\boldsymbol{\lambda}_T$ & $\boldsymbol{\lambda}_{inj}$ & Match approx. & $M(P_{10})$ \\ 
  \hline
  3.5 PN (2D) & 3.5 PN & 3.5 PN & $0.989 \pm 7.6\times10^{-4}$ \\
  3.5 PN (2D) & 3.5 PN & 0 PN & $0.989 \pm 8.7\times10^{-4}$ \\
  0 PN (1D) & 3.5 PN & 0 PN & $0.985 \pm 1.5\times10^{-3}$ \\ 
\end{tabular}
\caption{\label{tab:bank_eff_runs} A summary of the bank efficiency calculations done. The first column indicates the PN order of the template waveforms, and in parenthesis the dimensionality of the template banks. The second column similarly shows the PN order of the injected waveforms, highlighting that we have always used 3.5 PN injection waveforms. The third column shows the PN order used in the calculation of the match between template and injected waveforms. The last column shows the mean $10^{\mathrm{th}}$ percentile match across all 181 \hbbh\ sources. The one standard deviation error is quoted.}
\end{table}

\section{Visibility of stellar-mass BBH in {\em LISA}}
\label{sec:visibility lisa}

We now calculate, for template banks of $\mathcal{O}(10^2) - \mathcal{O}(10^5)$ templates, the minimal signal-to-noise ratio required to claim a detection in a matched filter search. We first assume that we have a segment of data consisting only of noise, $d(t) = n(t)$. If we filter this data with one template, the SNR would be a random value which, for a Gaussian background, follows the Rayleigh distribution \cite{Moore:2019pke}: 
\begin{equation}
    p(\rho) = \rho \exp{\left(\frac{-\rho^2}{2}\right)}.
\end{equation}
The probability of obtaining a value of $\rho$ greater than some threshold $\rho_{\rm thr}$ can be computed by integrating the distribution above the threshold, 
\begin{equation}
    p(\rho > \rho_{\rm thr}) = \int_{\rho_{\rm thr}}^{\infty} \rho \exp{\left(\frac{-\rho^2}{2}\right)}\, \mathrm{d}\rho = \exp{\left(\frac{-\rho_{\rm thr}^2}{2}\right)}.
    \label{eq:snr_integral}
\end{equation}
We would like to find the value of $\rho_{\rm thr}$ above which we can confidently claim a detection. The false alarm probability (FAP) is then the chance of finding $\rho > \rho_{\rm thr}$ in the noise hypothesis. We can choose an acceptable value of FAP, say $10^{-2}$ and use Eq. \eqref{eq:snr_integral} to solve for $\rho_{\rm thr}$: 
\begin{equation}
    \rho_{\rm thr} = \sqrt{-2 \ln{\mathrm{FAP}}}.
    \label{eq:snr_threshold}
\end{equation}
Therefore, with $\mathrm{FAP} = 10^{-2}$ the threshold is $\rho_{\rm thr} \approx 3.0$. However, the above discussion assumes that we filter the data with only one template. With $N_T$ templates, the FAP is the probability that at least one template has $\rho > \rho_{\rm thr}$. First, consider the probability that none of the $N_T$ templates have $\rho > \rho_{\rm thr}$, that is $(1 - p(\rho > \rho_{\rm thr}))^{N_T}$. The FAP is the compliment of this, or 

\begin{equation}
    \mathrm{FAP} = 1 - (1-p)^{N_T}.
\end{equation}

 For very small $p$, we can approximate this as $\mathrm{FAP} \approx 1 - ( 1 - p \cdot N_T) = p(\rho > \rho_{\rm thr}) \cdot N_T$. We see that the number of templates becomes a trials factor on the false alarm probability. We must then scale down our required FAP by the number of templates in Eq. \eqref{eq:snr_threshold}, 

\begin{equation}
    \rho_{\rm thr} = \sqrt{-2 \ln{\frac{\mathrm{FAP}}{N_T}}}.
\end{equation}

Finally, using $\mathrm{FAP} = 10^{-2}$ and $N_T = 10^2 - 10^5$, we find a threshold SNR of $\rho_{\rm thr} = 4.3$--$5.7$. This is a significant improvement over the previously quoted $\rho_{\rm thr} = 14$ \cite{Moore:2019pke} or $\rho_{\rm thr}=8$ quoted in Ref.\,\cite{Wong:2018uwb}. The lower threshold further improves the feasibility of archival matched-filter searches in {\em LISA}, increasing the number of detectable stellar mass binary black holes in {\em LISA} by a factor of $\sim 6.$

\section{Conclusions and Future Directions}
\label{sec:conclusions}
The simultaneous operation of third generation ground-based gravitational-wave observatories, such as the Einstein Telescope and Cosmic Explorer, and a space-based detector, {\em LISA}, provides a unique opportunity for multiband observations of stellar-mass and intermediate-mass binary black hole inspirals and mergers. Such multiband observations greatly improve the bounds one can place on general relativity and alternative theories of gravity \cite{Barausse:2016eii, Carson:2019kkh, Gupta:2020lxa, Datta:2020vcj}. However, the computational cost required to dig  the signals out of {\em LISA} data in a blind search would be formidable, requiring signal-to-noise ratios greater than 14 to make confident detections. 

We have shown that 3G observatories would constrain most of the signal parameters---the component masses, spins, sky position, and time of coalescence---so tightly that it would be possible to carry out the search for these signals over a vastly reduced parameter space in {\em LISA} data. Indeed, high fidelity measurements enabled by the 3G observatories considered in this paper imply that archival searches in {\em LISA} data require only a one-dimensional template bank over the binary's chirp mass---the only parameter that is measured better by {\em LISA} than 3G observatories. Thus, the volume of parameter space necessary to search over is greatly reduced. With template banks containing only $10^2$--$10^5$ templates, archival searches in {\em LISA} are feasible for signals of SNR as low as $\sim 4$--$6.$ This would allow for the possibility of $\mathcal{O}(100)$ multiband gravitational-wave detections per year, a number that can vastly improve the quality of tests of general relativity that can be performed by 3G observatories or {\em LISA} by themselves \cite{Gupta:2020lxa, Datta:2020vcj}.

We have made several implicit approximations in this study that would need to be reexamined to ascertain that the principal conclusions of this paper remain valid. Firstly, we have assumed that the observed binaries would have negligible residual eccentricity when their signals enter the {\em LISA} sensitivity band. This is true for most binaries that spend millions of years to slowly spiral-in and merge during which radiation reaction tends to circularize the orbits. However, black hole binaries that form in rich clusters could have non-negligible eccentricities when {\em LISA} observes them. 3G observatories should be able to constrain residual eccentricities to a pretty good accuracy which could then be used to reduce the search space. Even if the eccentricity in the audio band of 3G observatories is vanishingly small it will be possible to evolve the orbits back and limit the search parameter space in the {\em LISA} band. Eccentricity will likely be a parameter that would not be measured very well by 3G observatories (as any residual eccentricity would have largely decayed) and {\em LISA} will likely constrain it better.

Secondly, we have assumed that companion black hole spins are aligned with the orbital angular momentum of the system. Spin-orbit and spin-spin couplings will significantly alter the orbital evolution only when spins are large and misaligned with the orbit. This will be a relatively small part of the parameter space. Moreover, spin effects occur at higher PN orders and are measured better by 3G observatories than {\em LISA}. It is, however, important to confirm that precessional effects are negligible in archival searches especially since they are cumulative effects and the signals spend many more cycles in the {\em LISA} band than they do in the audio band.

Finally, if the binaries live in a gas-rich environment then the inertial drag could accelerate the rate at which the companions spiral in. Such drags might not be relevant when the binaries enter the audio band of 3G detectors and hence it would not be possible to correct for the presence of environment. In the same spirit, we have not included any ultra-light boson clouds or other fields that might surround the companion black holes in the low-frequency phase of the evolution. This could alter the orbital evolution but would not be relevant at later stages. 

While not all of these effects are equally important, it is necessary to investigate the cost of including them in an archival search as they could potentially reveal important mechanisms in play in the formation, evolution, and environments of stellar-mass and intermediate-mass black hole binaries. 

\section*{Acknowledgements}
We thank Anuradha Gupta for providing access to her simulation of binary black holes that can be observed both in {\em LISA} and the 3G network.  BE is supported by the Eberly Graduate Fellowship of Penn State. SS is supported by the Eberly Postdoctoral Fellowship of Penn State. BSS and SB are supported in part by NSF Grant No. PHY-1836779. This paper has the LIGO document number LIGO-P20xxxxx.
\bibliography{references}
\end{document}